\documentclass[fleqn,12pt]{wlscirep}
\usepackage[utf8]{inputenc}
\usepackage[T1]{fontenc}
\usepackage{subcaption}
\title{Musical Training, but not Mere Exposure to Music, Drives the Emergence of Chroma Equivalence in Artificial Neural Networks}

\author[1,*]{Lukas Grasse}
\author[1]{Matthew S. Tata}

\affil[*]{lukas.grasse@uleth.ca}

\affil[1]{Canadian Centre for Behavioural Neuroscience, Department of Neuroscience, University of Lethbridge, AB, Canada, T1K 3Y4}

\begin{abstract}
Pitch is a fundamental aspect of auditory perception. Pitch perception is commonly described across two perceptual dimensions: \textit{pitch height} is the sense that tones with varying frequencies seem to be higher or lower, and chroma equivalence is the cyclical similarity of notes octaves, corresponding to a doubling of fundamental frequency. Existing research is divided on whether chroma equivalence is a learned percept that varies according to musical experience and culture, or is an innate percept that develops automatically. Building on a recent framework that proposes to use ANNs to ask 'why' questions about the brain, we evaluated recent auditory ANNs using representational similarity analysis to test the emergence of pitch height and chroma equivalence in their learned representations. Additionally, we fine-tuned two models, Wav2Vec 2.0 and Data2Vec, on a self-supervised learning task using speech and music, and a supervised music transcription task. We found that all models exhibited varying degrees of pitch height representation, but that only models trained on the supervised music transcription task exhibited chroma equivalence. Mere exposure to music through self-supervised learning was not sufficient for chroma equivalence to emerge.  This supports the view that chroma equivalence is a higher-order cognitive computation that emerges to support the specific task of music perception, distinct from other auditory perception such as speech listening. This work also highlights the usefulness of ANNs for probing the developmental conditions that give rise to perceptual representations in humans. 

\end{abstract}

\begin{document}

\flushbottom
\maketitle

\fancypagestyle{firstpage}{%
  \chead{\sffamily\small Manuscript Under Review}
  \rfoot{}
}

\chead{\sffamily\small Manuscript Under Review}

\setlength{\headheight}{14pt}

\thispagestyle{firstpage}

\section{Introduction}

Auditory perception spans a variety of dimensions, such as pitch and loudness. Although sound is initially decomposed into component frequencies via the basilar membrane and auditory nerve, the perception of pitch is more abstract, with the perceptual dimensions of \textit{pitch height} and \textit{chroma} varying  independently.  Pitch height is the sense of absolute increase or decrease of (usually) the fundamental frequency of tones, and the percept of rising or falling pitch is sometimes called \textit{spectral listening}.  Chroma refers to a cyclical percept as fundamental frequencies double; this is the sense that notes separated by octaves sound similar.  Pitch is therefore commonly represented by a multidimensional helix \cite{shepard1964circularity,shepard1982geometrical}, with pitch height represented vertically, and chroma represented angularly. Thus, points on the spiral that are vertically aligned are said to exhibit \textit{chroma equivalence} despite having different heights.  Although the perception of pitch height is relatively well understood and consistent, there is a debate in the literature as to whether chroma equivalence is learned and varies due to factors such as culture and musical experience, or is an innate percept shaped by biological or evolutionary constraints. For example, early research such as \cite{allen1967octave} found differences between musicians and non-musicians, with musicians rating the same notes from different octaves as more similar relative to non-musicians. However, later research such as \cite{ueda1987perceptual} did not find a significant correlation with musical ability. 

\begin{figure}[!htbp]
  \centering
    \includegraphics[width=1.0\textwidth]{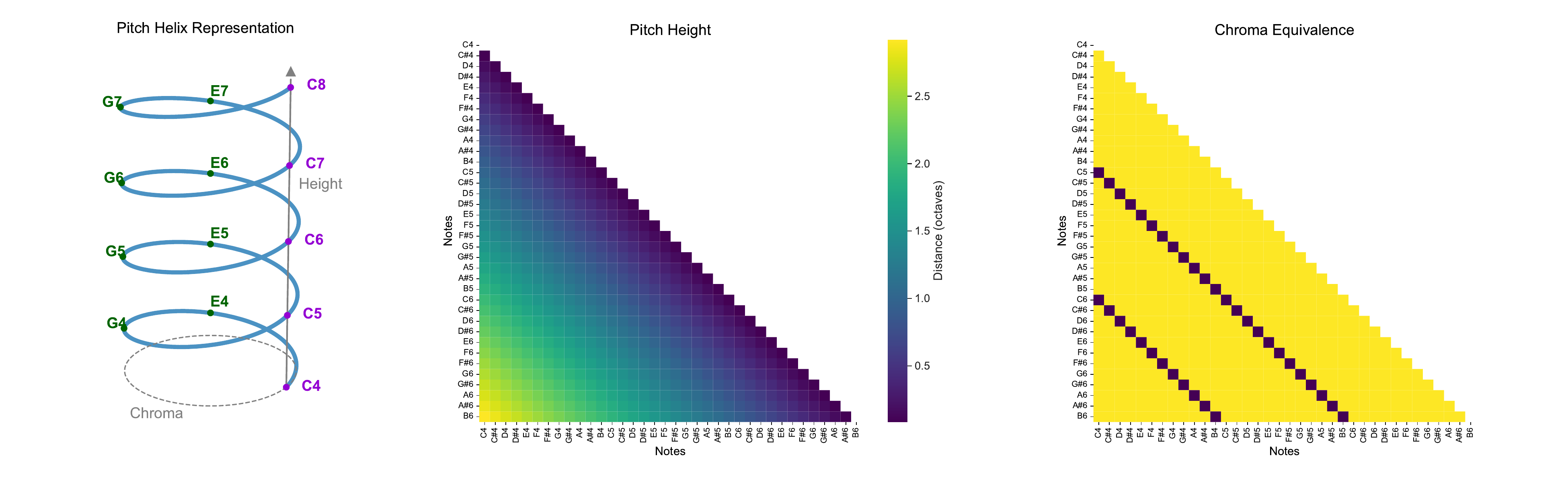}
    \caption{\textbf{Representations of Pitch}.  Helical Representation of Pitch.The vertical dimension corresponds to absolute pitch height as distance in fundamental frequency. The circular horizontal dimension corresponds to the repeating cyclical octaves of chroma. }
    \label{fig:representations}
\end{figure}

Neuroimaging has contributed some insight to this debate: research has found that pitch height and chroma are handled by distinct areas in the auditory cortex and with different time courses. For example \cite{warren2003separating} used fMRI to show that pitch height information corresponded to activity in the planum temporale, while chroma information corresponded to activation of the planum polare. A study using EEG  \cite{regev2019evidence} to investigate the mismatch negativity (MMN) found that an MMN was only elicited for differences in pitch height, but not for differences in chroma, even in participants who could perceive chroma differences in a perceptual task. This points to chroma potentially being a higher-order cognitive process relative to the more automatic processing of pitch height. A recent study of non-musicians using MEG \cite{chang2025temporally}, found that pitch height and chroma information could be decoded at distinct latencies over a few hundred milliseconds post-stimulus onset, with pitch height information being decoded first, then a brief window in which chroma information could be decoded, followed again by pitch height. The decoding of pitch height was earlier and more robust, which supports the idea that the processing of pitch height is automatic, while perception of chroma is secondarily computed. Since chroma equivalence could only be briefly decoded and the task was not specific to chroma equivalence, the study adds evidence that pitch height and chroma equivalence are independent features of auditory perception that are processed automatically even in the absence of a relevant task. Additionally, since the study participants were non-musicians it is clear that chroma equivalence is not necessarily dependent on musical training, although it may reflect a computation that underlies music perception and occurs in all (or most) individuals who have listened to music. 

This is in contrast to \cite{jacoby2019universal}, which conducted a cross-cultural study investigating chroma/octave equivalence using a singing reproduction task in which participants heard tone sequences in different registers and sang them back. The participants were either from the United States or from an Amazonian tribe that had been relatively isolated from exposure to western music. They found that US participants tended to sing the sequences back at an integer multiple of the original fundamental, whereas the Amazonian tribe did not. That is, US participants tended to replicate the original chroma.  They also found that musicians among the US participants exhibited stronger chroma equivalence than non-musicians.  These results suggest that chroma equivalence might be a learned cultural percept rather than an innate or fundamental one. 

Previous research with infants has investigated whether chroma equivalence emerges early in development. Using a habituation-dishabituation procedure, \cite{demany1984perceptual} found that three-month old infants are sensitive to both pitch height and chroma, suggesting that chroma perception does not depend on musical experience. Likewise, a recent EEG study on infants \cite{gennari2025neural} also demonstrated that pitch height and chroma could be decoded from EEG, with pitch height decoding performance better earlier in the trial and chroma decoding performing better at later latencies. This research on infants is interesting in that it has also been shown that infants can outperform adults on pitch discrimination \cite{lau2021infant}, with 3 and 7 month old infants performing at the level of trained musicians and outperforming non-musicians. 

Psychophysical, cross-cultural, and neuroimaging evidence therefore indicate that pitch and chroma-equivalence are processed by distinct mechanisms; however, it is unclear whether chroma equivalence emerges automatically as a consequence of fundamental auditory development in general, or if it follows from a listener's particular experiences and interactions with tonal stimuli during music listening.  

A recently proposed framework \cite{kanwisher2023using} demonstrates that artificial neural networks (ANNs) can be powerful tools to ask 'why' questions of minds and brains. The idea is that if a perceptual effect or illusion is exhibited by ANNs under only certain specific conditions, then it may point to those conditions being important for emergence of that process in human brain development. The framework outlines that the four main dimensions across which ANN training can be modulated, and thus tested, are by modulating the training data, network architecture, learning algorithm, or objective function. Chroma equivalence is an auditory perceptual quality with poorly understood factors that give rise to its development. This makes ANNs a strong choice for modeling chroma equivalence, because we can train them under controlled conditions to see what it takes for chroma equivalence to emerge. 

Existing research has recently studied pitch perception using ANNs.   \cite{ahmad2016harmonic} trained a biologically plausible architecture of pitch neurons and found pitch-perception effects evoked by iterated ripple noise stimuli, which elicit the percept of pitch in noise for human listeners. Furthermore, their model exhibited a well-known perceptual effect called the illusion of the Missing Fundamental, suggesting chromatic perception of complex tones. Similarly, \cite{saddler2021deep} trained convolutional neural networks to predict the fundamental frequency of stimuli and then tested the networks on various illusions and perceptual effects. These papers both trained supervised models to directly estimate the pitch of stimuli during training.  \cite{tuckute2023many} took the approach of testing deep ANNs that were not explicitly trained to perform pitch estimation, but were rather trained using self-supervised learning or supervised learning for different auditory tasks. To evaluate the similarity of the ANNs to the brain, they then fit linear mappings from the neural activations of the ANNs to recorded fMRI data from participants presented with the same stimuli from a natural-sound listening task, and also used Representational Similarity Analysis (RSA) \cite{nili2014toolbox} to compare the neural activations to ANN model activations, showing similarities in the encoding of pitch features between the ANN and human brains.

To our knowledge, no prior studies have examined the emergence of chroma equivalence in ANNs. We therefore set out to test a set of related questions using supervised and unsupervised ANNs:  First, do either pitch height or chroma equivalence (or both) emerge after self-supervised training alone?  In other words, is mere exposure to speech or music in the training data sufficient for pitch height and chroma equivalence to emerge. Second, is music necessary in the training data for these percepts to emerge, or is any complex sound, such as speech or environmental noise sufficient?  Third, if self-supervised networks fail to encode either dimension, despite having music in their training data, does supervised fine-tuning on a music task drive emergence of pitch and/or chroma equivalence?

\section{Methods}
We evaluated auditory artificial neural networks trained on various supervised and self-supervised tasks to test whether their learned representations exhibited spectral listening, chroma equivalence, or a mixture of both. Additionally, we fine-tuned self-supervised speech models with a combination of speech and music using self-supervised learning. We then fine-tuned the models to perform polyphonic music transcription to test the effect of learning an active music task relative to mere music exposure. The following sections outline the stimuli, neural network architectures, training, and analyses. 

\subsection{Stimuli for Spectral and Chroma Listening}
For evaluation of spectral vs. chroma listening we selected notes from the NSynth dataset \cite{nsynth2017}, a large annotated dataset of synthesized musical notes with unique pitch, timbre, and envelopes. We first filtered the training set to only include instruments that contained all notes for octaves 4-6, and then selected a balanced subset of 30 instruments consisting of 10 flutes, 10 guitars, and 10 keyboards. We used notes with a midi velocity of 100 for evaluation. 

\subsection{Representational Similarity Analysis}

To evaluate the degree of pitch height and chroma representation present in the learned ANN embeddings, we performed representational similarity analysis (RSA) \cite{nili2014toolbox}. RSA works by creating a representational dissimilarity matrix (RDM) for a set of stimuli that is the pairwise dissimilarity for each stimulus to all other stimuli. These RDMs can be computed from a variety of sources such as brain activity recordings from fMRI, MEG, EEG, and also from neural activations from the hidden layers of ANNs.  Importantly, hypothesized perceptual models also imply patterns in RDMs.  For example, a listener who hears only pitch height but not chroma would be represented by an RDM such as the one shown in the middle panel of Figure  \ref{fig:representations}.  Likewise, a listener who encodes only chroma equivalence could be represented by the RDM in the left panel of Figure \ref{fig:representations}.   RSA then directly compares these RDMs by simple correlation to ask questions such as which ANN model RDM best matches the RDM of another dataset, or another model. In this research, we calculated RDMs for a variety of ANNs and asked how well they are explained by a pitch height model that corresponds to purely spectral listening, or a chroma equivalence model. Individual RDMs were recorded for each instrument for a total of 30 RDMs per ANN. RDMs were calculated using correlation distance which is defined as $1-|r|$ where $r$ is the Pearson correlation coefficient. This correlation distance was calculated for each pair of note embeddings to create an RDM for each network/instrument combination. RDMs were then compared to the pitch height and chroma equivalence models using Spearman Rank Correlation. We also compute the noise ceiling for each model using the individual instrument RDMs \cite{nili2014toolbox}. The noise ceiling provides an estimate of the maximum range that an ideal model would achieve if it perfectly predicted the network RDMs. The lower bound of the ceiling is computed by correlating each RDM with all other RDMs in a leave-one-out approach, and the upper bound is calculated by correlating each RDM with all other RDMs including itself. For RSA analysis, we used the Python RSA Toolbox \cite{van2025python}.

\subsection{Neural Network Architectures and Training} 
We evaluated several neural network architectures and baseline "hard-coded" representations. The key properties of each model are summarized in Table \ref{tab:models}. All networks use the transformer architecture \cite{vaswani2017attention}, and self-supervised learning (SSL), supervised learning (SL), or self-supervised learning + supervised fine-tuning (SSL+SFT). The Hugging Face \cite{wolf2019huggingface} implementation of each model was used.

To test the hypothesis that musical exposure in the training data would lead to increased chroma equivalence perception, we fine-tuned Wav2Vec 2.0 \cite{baevski2020wav2vec} and Data2Vec \cite{baevski2022data2vec} models on a speech and music data set. Both models were originally trained solely on the Librispeech dataset \cite{panayotov2015librispeech} which contains only speech, and we further fine-tuned them using self-supervised learning on 460 hours of speech from Librispeech and 460 hours of music from the LAION-DISCO-12M music dataset \cite{laion2024disco}. Both networks were fine-tuned for 50000 steps on the speech/music dataset using the official implementations contained in the fairseq toolkit \cite{ott2019fairseq}. Networks were trained on a machine containing two Nvidia 4090 GPUs, using gradient accumulation to emulate the number of GPUs used in the original training of Wav2Vec 2.0 and Data2Vec, which were 64 and 16 respectively. All other hyperparameters were kept the same as the original configuration in the fairseq toolkit. The checkpoints were then converted to Hugging Face format for evaluation and further fine-tuning on a supervised music transcription task.

\begin{table}[!htbp]

\footnotesize\sf\centering
\caption{\textbf{Summary of neural network architectures and baseline models evaluated.}}
\label{tab:models}
\begin{tabular}{l l l l l}
\toprule
\textbf{Model} & \textbf{Variant} & 
\textbf{Paradigm} & \textbf{Training Task} & \textbf{Training Data} \\
\midrule

Wav2Vec 2.0 \cite{baevski2020wav2vec} & wav2vec2-base & SSL & Contrastive loss (masked) & Librispeech \\
Data2Vec \cite{baevski2022data2vec} & data2vec-audio-base & SSL & Student-teacher (masked) & Librispeech  \\
Whisper \cite{radford2023robust} & whisper-base & Supervised & Multitask ASR & Multilingual speech  \\
MERT \cite{li2023mert} & MERT-v1-95M & SSL & Student-teacher (Acoustic+CQT) & Music \\
AST \cite{gong2021ast} &  ast-finetuned-audioset & SSL+SFT & Audio classification & ImageNet + Audioset \\
\midrule

Spectrogram & 128 Bands & Transform & Mel scale & N/A  \\
CQT \cite{brown1991calculation} & 336 Bins & Transform & Constant-Q & N/A \\
Cochleagram \cite{chang2025temporally} & 128 filters & Transform & ERB Gammatone & N/A \\
\bottomrule
\end{tabular}
\end{table}

\subsection{Music Transcription Finetuning}

To test the hypothesis that supervised learning of an explicit music task would lead to increased representation of chroma equivalence in ANNs, we fine-tuned Wav2Vec 2.0 and Data2Vec to perform the task of transcribing polyphonic piano music. To do this, we used the MAESTRO dataset \cite{hawthorne2018enabling}, a dataset consisting of 200 hours of paired audio and MIDI data from an international piano competition. For training, we divided the dataset into training/test subsets with a 90\%/10\% split. Following the training approach used in a recent architecture for polyphonic music transcription \cite{bittner2022lightweight}, we trained the model to predict which MIDI notes were active for each time step of the encoder output. Training was implemented using Binary Crossentropy Loss and the Wav2Vec2ForAudioFrameClassification and Data2VecAudioForAudioFrameClassification implementations from the Hugging Face Transformers library for Wav2Vec 2.0 and Data2Vec respectively. The training task is highly unbalanced as most notes are not active for a given audio frame, so following \cite{bittner2022lightweight}, we weighted the positive class with 0.95 and negative class with 0.05. MIDI notes were mapped to the range 0-88 corresponding to the 88 piano keys, and audio/midi sequences were divided into 2-second chunks. We trained each model for 10 epochs of the dataset with a learning rate of \textit{3e-5} and a batch size of 16. Importantly, none of these models were trained to perform chroma classification.

\section{Results}

\begin{figure}[!htbp]
    \centering
    \begin{subfigure}[b]{1.0\textwidth}
        \centering
        \caption{\textbf{Absence of Chroma Equivalence in Pre-trained Models.} Pretrained models encode pitch height but do not exhibit chroma equivalence regardless of whether they have music in their initial training data.}
        \includegraphics[width=0.6\textwidth]{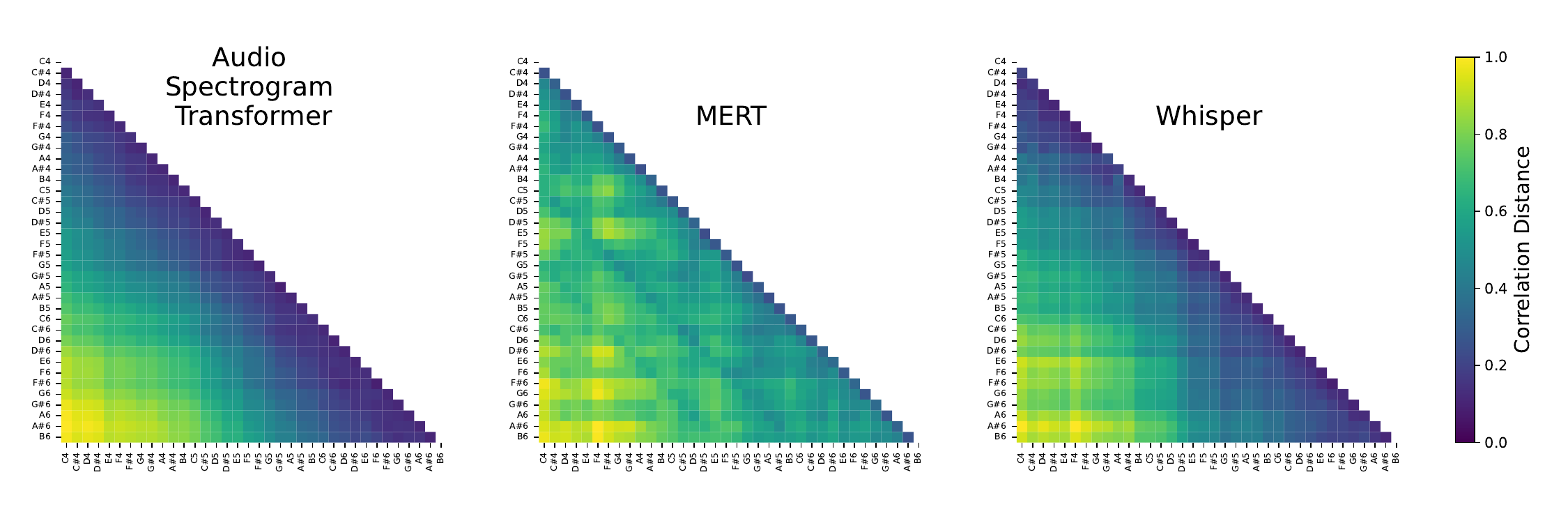}
        \label{fig:pretrained_model_rdms}
    \end{subfigure}

    \begin{subfigure}[b]{1.0\textwidth}
        \centering
        \caption{\textbf{Chroma Equivalence in Hard-Coded Models} The CQT model does exhibit chroma equivalence, but it is designed to do so.}
        \includegraphics[width=0.6\textwidth]{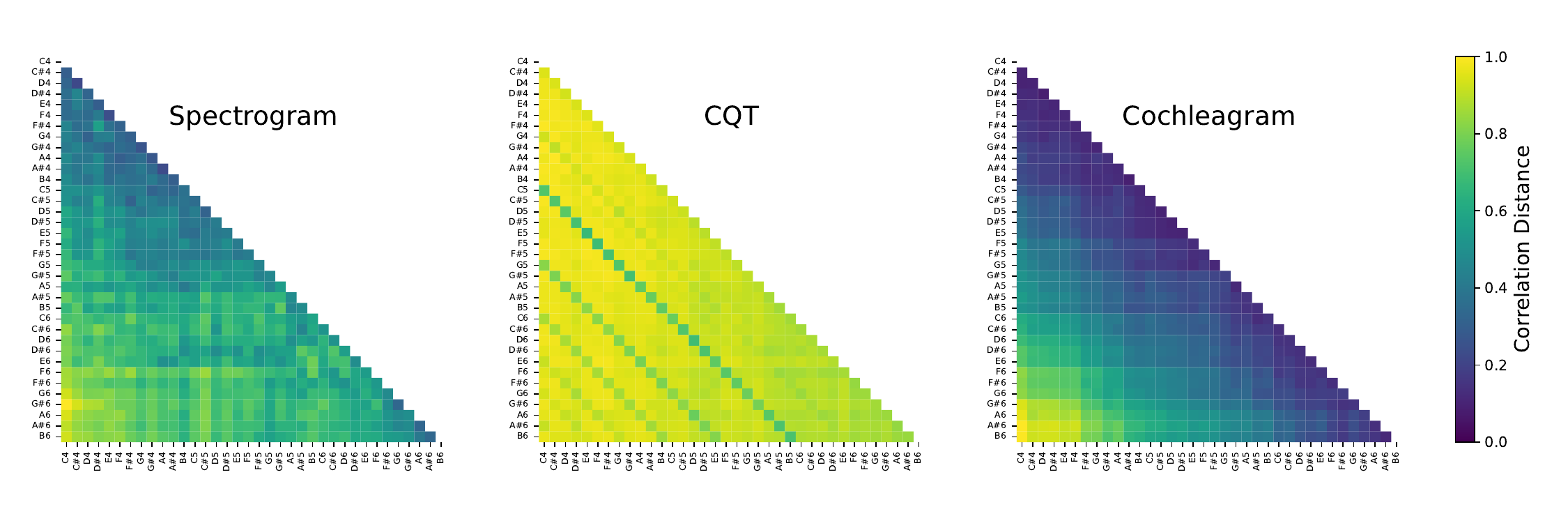}
         \label{fig:hardcoded_model_rdms}
    \end{subfigure}

    \begin{subfigure}[b]{1.0\textwidth}
        \centering
        \caption{\textbf{Inclusion of Music in Training Data} Incorporating music data into Wav2vec 2.0 and Data2vec by self-supervised fine tuning was not sufficient for chroma equivalence to emerge.}
        \includegraphics[width=0.8\textwidth]{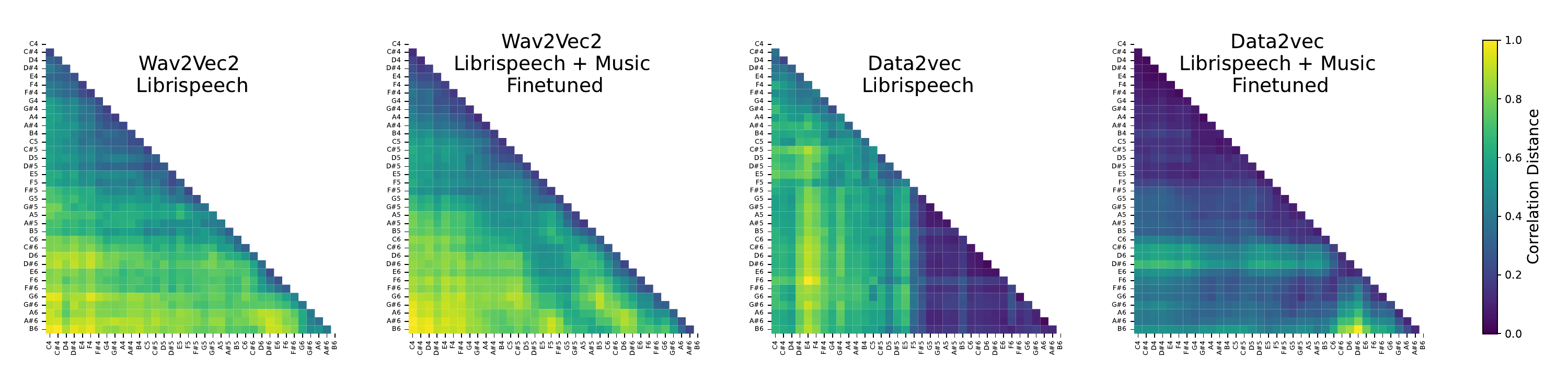}
        \label{fig:ssl_finetune_model_rdms}
    \end{subfigure}

    \begin{subfigure}[b]{1.0\textwidth}
        \centering
        \caption{\textbf{Music Task Fine-Tuning} Further fine-tuning of Wav2vec 2.0 and Data2vec on an explicit musical task (note transcription) leads to emergence of chroma equivalence.}
        \includegraphics[width=0.8\textwidth]{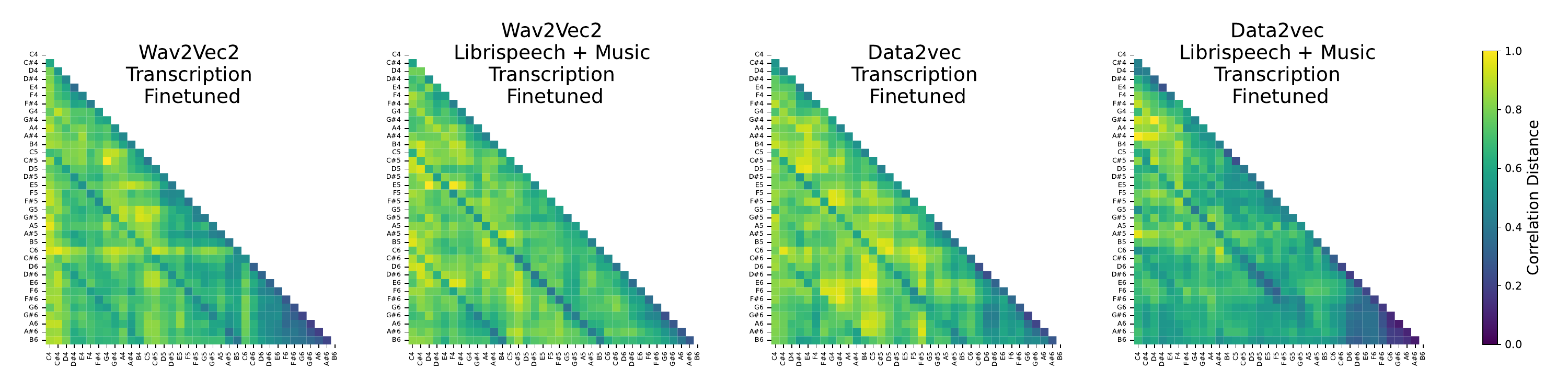}
        \label{fig:transcription_finetune_model_rdms}
    \end{subfigure}

        \caption{\textbf{Note Representational Dissimilarity Matrix (RDM) Plots.} RDM plots for NSynth test note dataset. RDMs were averaged across 30 instruments and normalized for plotting.   }
         \label{fig:model_rdms} 
\end{figure}

\begin{figure}[!htbp]
    \centering
    
    \begin{subfigure}[b]{1.0\textwidth}
         \centering
        \caption{Pretrained Models}
         \includegraphics[width=1.0\textwidth]{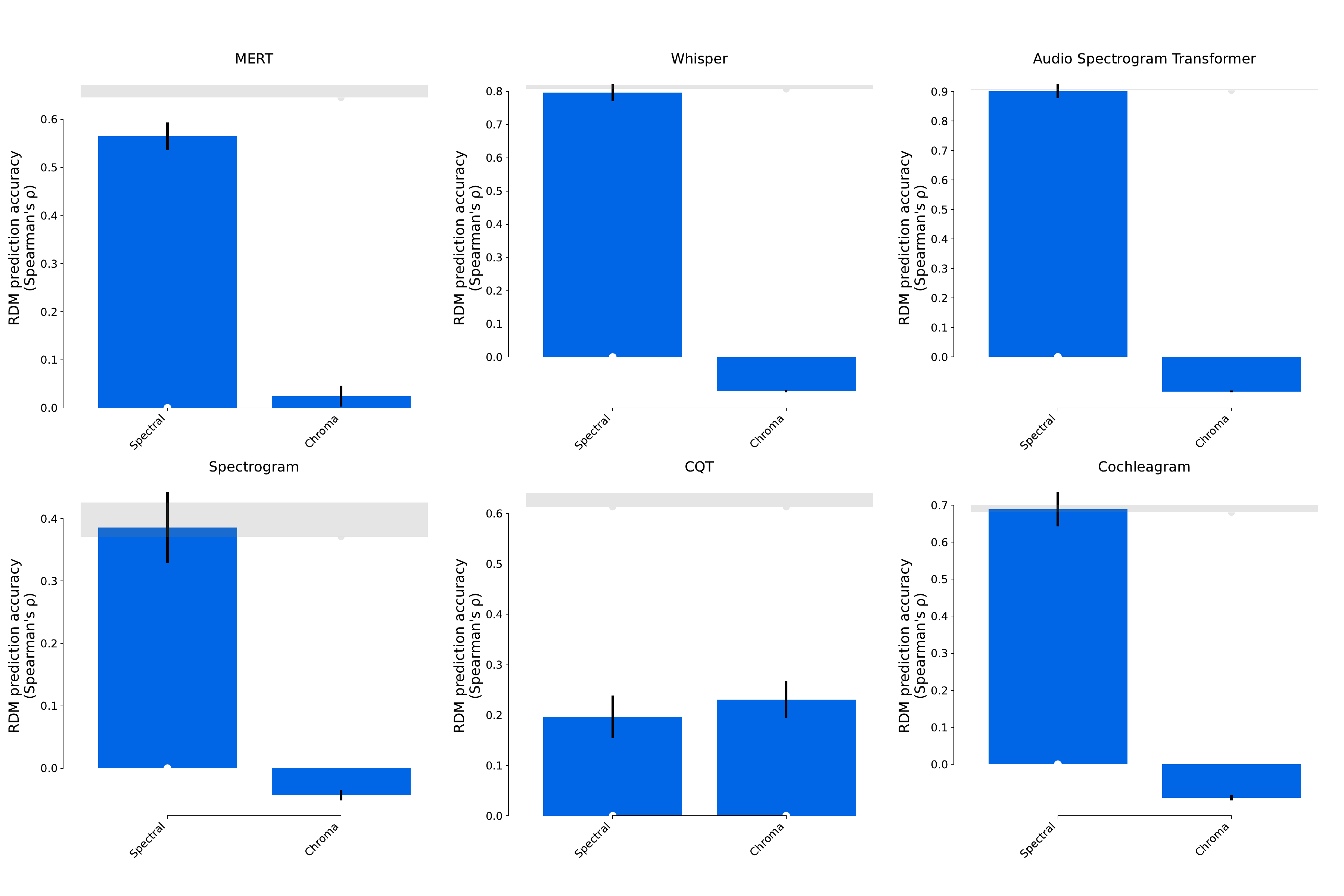}

         \label{fig:pretrained_models} 
    \end{subfigure}

    \begin{subfigure}[b]{1.0\textwidth}
         \centering
         \caption{Finetuned Models}
         \includegraphics[width=1.0\textwidth]{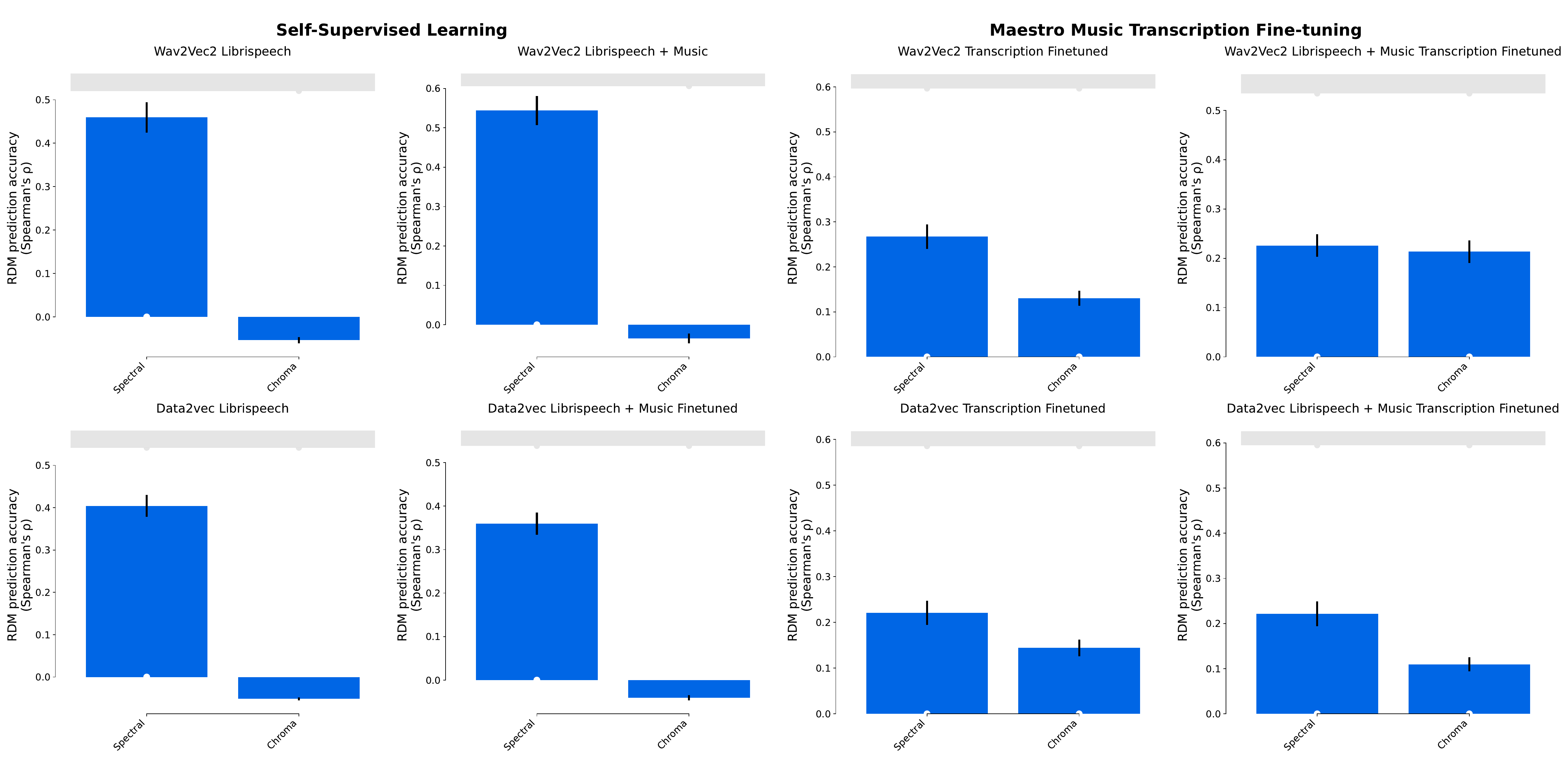}

         \label{fig:finetuned_models} 
    \end{subfigure}
    
    \caption{RSA Analysis for Pretrained and Finetuned Models.}
    \label{fig:combined_analysis}
\end{figure}

Representational Dissimilarity Matrices (RDMs) were calculated for all networks and hard-coded baselines. Plots of the RDMs averaged across all instruments are shown in Figure \ref{fig:model_rdms}. All ANNs appear to exhibit some degree of pitch height representation, regardless of learning task and training on speech, music, or AudioSet. Next, we used RSA to quantify how well the network RDMs matched the Pitch Height and Chroma Equivalence Model RDMs, as shown in Figure \ref{fig:representations}. Model RDMs were compared to ANN RDMs using Spearman's correlation coefficient. The grey bar on each plot indicates the noise ceiling. The half-moon indicators on each bar indicate significant differences from zero, and dewdrop indicators on the noise ceiling bar indicate significant differences from the noise ceiling. The error bars show the standard error of the mean (SEM). Statistical tests were performed using t-tests with $\alpha=0.01$ and corrected for multiple comparisons using Bonferroni correction. As shown in Figure \ref{fig:combined_analysis}, all of the self-supervised pretrained ANNs exhibited strong pitch height encoding, but none of them exhibited chroma equivalence, regardless of their training algorithm and training data. Some hard-coded baseline models exhibited chroma equivalence, namely the CQT transform, which is expected since the CQT is designed to represent auditory features according to their correspondence with the western musical scale. 

Figure \ref{fig:finetuned_models} shows the results of incorporating musical stimuli into the training sets of Wav2Vec 2.0 and Data2Vec using self-supervised fine tuning.  This incorporation of music into the training data was not sufficient for chroma equivalence to emerge. Following incorporation of music, Wav2Vec 2.0 did show an increase in pitch height representation relative to the version trained only on speech. This improvement did not appear to occur in Data2Vec. Figure \ref{fig:finetuned_models} also shows the effect of supervised fine-tuning on the music transcription task, which does elicit chroma equivalence in Wav2Vec 2.0 and Data2Vec. To test whether this effect was due to supervised fine-tuning in general, and not specific to a music related task, we also evaluated Wav2Vec 2.0 and Data2Vec using their respective checkpoints that were fine tuned on LibriSpeech to perform a speech recognition task.  In contrast to fine tuning on the music transcription task,  fine tuning on speech recognition yielded no increase in chroma equivalence and similar pitch height encoding for Wav2Vec 2.0, with a slight increase in pitch height encoding for Data2Vec.  Thus self-supervised learning on speech, music, and a mixture was sufficient for pitch height representation to emerge in the embeddings of these ANNs, but was not sufficient to yield chroma equivalence in those embeddings.  Likewise, supervised fine tuning on a speech recognition task also did not lead to chroma equivalence.  Supervised fine tuning on a music task, however, did cause chroma equivalence to emerge in these ANN models.

\section{Discussion}

Auditory perception of pitch has traditionally been represented with a helical representation comprised of pitch height and chroma (Figure \ref{fig:model_rdms})\cite{shepard1964circularity,shepard1982geometrical}. Pitch height refers to a perceived absolute increase or decrease in the fundamental frequency of a tone, whereas chroma refers to a cyclical perception of notes across octaves being perceived as similar. This dimension of chroma has been the subject of considerable debate in the literature regarding whether it is an innate function of auditory development, or whether it is acquired based on cultural and musical experience. Early research found differences between musical and non-musical individuals \cite{allen1967octave}, but this has not always been replicated \cite{ueda1987perceptual}.
Neuroimaging studies have also illuminated aspects of how chroma is represented in the brain.  The observation that pitch height and chroma information can be decoded from different locations and over different time scales \cite{warren2003separating,chang2025temporally} suggests distinct encoding mechanisms. Other evidence points to the possibility that pitch height might be automatic and universal, whereas chroma might be a higher-order cognitive process \cite{regev2019evidence}. This interpretation also suggests that chroma is a learned cultural phenomenon \cite{jacoby2019universal} that is not universal across all cultures and levels of musical experience.

The developmental origin of chroma perception is difficult to unravel due to the variability of both exposure to music and active musical experience, both from individual to individual and also across cultures\cite{jacoby2019universal}.  Although relatively few humans are trained musicians, virtually all humans have experienced musical listening at some point in their lives.  ANNs, by contrast, learn only from the data in their training sets.  An ANN trained only on speech has never "heard" music, and vice versa.   The present study sought to inform this question by controlling the degree to which music is present in the training data of ANN models, and asking whether mere inclusion of music or an explicit musical task were critical in the emergence of chroma perception.  We thus applied a recent framework proposed by \cite{kanwisher2023using} to test ANNs trained under different conditions to see which conditions elicited the representation of chroma equivalence. 

Importantly, this framework not only enabled us to ask whether musical experience would give rise to chroma equivalence, but also enabled us to differentiate the effect of passive music exposure vs. active musical learning. We accomplished this by fine-tuning recent self-supervised audio architectures on a dataset of speech and music that corresponded to passive music exposure, and also fine-tuning the models to perform the active supervised task of music transcription. We found that self-supervised models and supervised models trained on the non-musical task of speech recognition did not exhibit chroma equivalence in their learned neural representations. In contrast to this, both of the models that were fine-tuned to perform music transcription did exhibit chroma equivalence.  Our results suggest that pitch height is a relatively basic feature encoded by auditory systems that need to perform fundamental computations on the frequency structure of sound - regardless of whether those computations are performed by biological or artificial hearing systems.  Chroma equivalence, by contrast, is a percept that reflects some additional task or computation that the auditory system learns in service of music perception. We do not mean to imply that musical training \textit{per se} is required for chroma equivalence to emerge in auditory perception.  Clearly, many human listeners who are not musicians are nevertheless capable of perceiving chroma equivalence.  Rather, we suggest that the music transcription task that we used to fine tune Wav2Vec 2.0 and Data2vec required these networks to learn to represent chroma as a feature, and that human perception of music, as distinct from other complex sounds such as speech, likewise requires music-specific computations to be learned.  Chroma equivalence among humans might be a perceptual consequence of one such computation.   

However, we considered only two supervised tasks: speech recognition and musical note transcription. It is not clear whether other tasks that are not music-related might nevertheless give rise to chroma equivalence, and this is one limitation of our study. Audio Spectrogram Transformer performs supervised  audio classification, yet we found that it did not exhibit chroma equivalence. Interestingly, the musical student/teacher architecture of MERT is trained by supervised learning to recreate the CQT transform, which does not strictly yield chroma equivalence but rather seems to encode the consonances of the octave, fifth, and third intervals, which are the component notes of a major chord. This suggests that chroma equivalence might reflect a computation that is not necessarily specific to music, but does reflect chromatic processing more generally. 

One non-musical explanation for the emergence of chroma equivalence is that young children might learn chroma equivalence through vocal mimicry \cite{hoeschele2017animal}, as they must learn to recreate speech from their parents who speak in a different octave range than their own range. This hypothesis would provide a logical explanation for the previously described research into chroma equivalence in infants \cite{demany1984perceptual,gennari2025neural,lau2021infant}. Existing research has also considered vocal mimicry in other species, such as budgerigars \cite{wagner2019octave}, and found that they did not exhibit chroma equivalence, which led them to conclude that vocal mimicry was not linked to chroma equivalence. Future work training ANNs to perform vocal mimicry could test this developmental hypothesis more directly.  Interestingly, the RDM of MERT's embeddings (along with the CQT on which it is trained) hint not at chroma equivalence but rather at a model of tonal interval or chord structure: a small dissimilarity appears along the diagonal representing the interval of the perfect fifth (and also the third, in CQT), perhaps reflecting overlapping harmonics or representing some encoding related to musical consonance.  Future work could use RDM representations of consonance and dissonance to investigate whether music-trained networks encode human-like musical relationships between notes within the octave.

An important aspect of the fine-tuning task is that we made use of the MAESTRO dataset \cite{hawthorne2018enabling}, which is a dataset of polyphonic piano music with corresponding midi sequences. This is an important aspect of the task for several reasons: piano music spans several octaves and the polyphonic nature of the music means that in order to perform the transcription task, the model needed to track more than one note occurring simultaneously. It may be that chroma equivalence emerges as a mechanism to parse a superposition of notes occurring across a wide range of frequencies.  A future study could investigate the specifics of active musical training by comparing monophonic music to polyphonic music across different frequency ranges to see if monophonic music is sufficient for chroma equivalence to emerge, and whether a wide frequency range is required. This is relevant to the previous cross-cultural research by \cite{jacoby2019universal}, which found that the instruments produced by an Amazonian tribe that did not perceive chroma equivalence during singing also had a narrower frequency range relative to western instruments.

In conclusion, this study investigated whether pitch height or chroma equivalence (or both) emerged in ANNs trained on various self-supervised and supervised tasks. We found that pitch height representations were common across all ANNs regardless of training data or task, whereas chroma equivalence only emerged in models trained on the supervised task of music transcription. This points to chroma equivalence as a higher-level cognitive process in service of specific computations associated with music perception, relative to the more automatic process of pitch height representation. 

\section{Funding Declaration}

The research was funded by an NSERC Canada Discovery Grant (No. \#05659) and a Government of Alberta Centre for Autonomous Systems in Strengthening Future Communities grant to MT and an NSERC CGS-D Scholarship to LG.

\section{Contribution Statement}
LG wrote the main manuscript text, designed and implemented the experiments and analysis, and prepared the figures, all under the supervision of MT.  MT contributed to the supervision of the project, writing, review, analysis, and project administration. All authors approved the final version of the submitted article.

\bibliography{bibliography}

\end{document}